\documentclass{JAC2003}
\usepackage{graphicx}
\usepackage{booktabs}

\begin{document}
\title{MEASUREMENT OF NEUTRAL PARTICLE CONTAMINATION IN THE MICE MUON BEAM\thanks{Work supported by US NSF grant No. PHY-0970178}}

\author{R. R. Fletcher\thanks{rflet001@ucr.edu}, L. Coney, G. Hanson, UC Riverside, Riverside, CA 92521, U.S.A.\\ for the MICE collaboration}

\maketitle

\begin{abstract}
The Muon Ionization Cooling Experiment (MICE) is being built at the ISIS proton synchrotron at Rutherford Appleton Laboratory (RAL)
to measure ionization cooling of a muon beam. During recent data-taking, it was determined that there is a significant background
contamination of neutral particles populating the MICE muon beam. This contamination creates unwanted triggers in MICE, thus
reducing the percentage of useful data taken during running. This paper describes the analysis done with time-of-flight detectors,
used to measure and identify the source of the contamination in both positive and negative muon beams.

\end{abstract}

\section{Introduction}

Two possibilities for the next generation of particle accelerators are a Neutrino Factory and a Muon Collider. A Neutrino
Factory will produce an intense, narrow beam of neutrinos from the decay of muons in a storage ring. This beam of muon
neutrinos ($\nu_{\mu}$) and electron anti-neutrinos ($\bar{\nu{_e}}$) could be used to study neutrino oscillations and leptonic CP violation.

Because muons are more massive than electrons, synchrotron radiation is reduced by a factor of about $6\times10^{-10}$ in a Muon
Collider compared with an electron-positron ($e^+e^-$) collider. This enables construction of high energy lepton colliders that
are much smaller and less expensive than e$^+$e$^-$ colliders of similar center-of-mass energy. Unlike proton colliders such
as the LHC, collisions in a Muon Collider take place between fundamental particles with the full center-of-mass energy available
in all interactions. This makes it ideal for precise studies of the Higgs boson and also extensions to the standard model such
as supersymmetry. Both a Neutrino Factory and Muon Collider require high intensity muon beams to reach interaction rate goals.
Cooling is essential in producing a muon beam of sufficient intensity.

The Muon Ionization Cooling Experiment (MICE) is an international collaboration of physicists hosted by the Rutherford
Appleton Laboratory (RAL), Oxfordshire, UK \cite{MICE}. The purpose of MICE is to build and test a section of cooling channel designed to
reduce transverse emittance in a muon beam by at least 10\%. Due to the short lifetime of muons, standard beam cooling
techniques can not be used. Ionization cooling provides an alternative method that can cool a beam of muons very quickly \cite{MICE}
\cite{coney}.

The cooling channel section is composed of liquid hydrogen cells to reduce momentum in all directions and then
radio frequency cavities to replace momentum in only the beamline direction.
\begin{figure}[ht]
   \centering
   \includegraphics*[width=80mm]{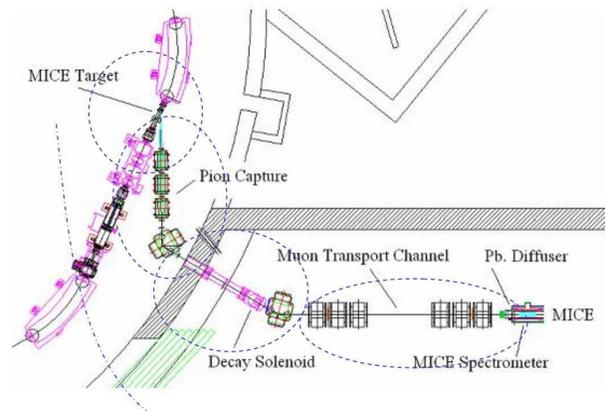}
   \caption{Top down drawing of the layout of the upstream beamline components in the ISIS vault and MICE experimental hall.}
   \label{schem}
\end{figure}
Measurements of emittance will be done by two scintillating fiber (SciFi) trackers located at the input and output of the
cooling channel. They will each be placed inside of 2-m long, 4-T superconducting solenoid magnets. SciFi trackers will
allow tracking of individual particles through the cooling channel \cite{coney}\cite{status}\cite{tracker}.

MICE is located at the ISIS proton synchrotron facility at RAL. The muon test beam is
produced by dipping a hollow titanium target into the ISIS proton beam. A quadrupole triplet captures pions produced in the
target that are then momentum selected by a dipole magnet. Following the dipole is a 5-T superconducting solenoid used to increase
the path length of the pions traveling though it, allowing time for them to decay. Muons from the decays are then momentum
selected by a second dipole and transported through the final upstream section by another two sets of quadrupole triplets (see Fig.\ref{schem})\cite{coney}\cite{status}.

The final upstream section also contains particle identification detectors (PID). Most measurements are made with a
set of time-of-flight detectors (TOF)\cite{tof}. TOF0, located just upstream of the last quadrupole triplet, is made of scintillating
bars with dual photomultiplier tube (PMT) readout, positioned in two planes to make transverse direction counter arrays.
TOF1 is placed just after the last quadrupole triplet and is of similar design to TOF0 \cite{coney}\cite{status}.

Data aquisition (DAQ) is controlled by setting one TOF as the trigger station. When a signal is registered in this station the DAQ
records the signal and any coincident signals in the other TOF. For example, if the trigger station is set to TOF1 and a particle
causes a hit in TOF0 and then TOF1, it will cause a trigger and the hits in both will be recorded in the DAQ. However, if the particle
causes a hit in TOF0 and then is scattered out of the beam or is stopped and does not reach TOF1 then the signal is not
recorded. The DAQ only allows triggers within a time window after the target is operated (spill gate) that for normal operation is set to 3.2 ms.

\section{DATA TAKING - SUMMER 2010}
\begin{figure}[htb]
   \centering
   \includegraphics*[width=70mm]{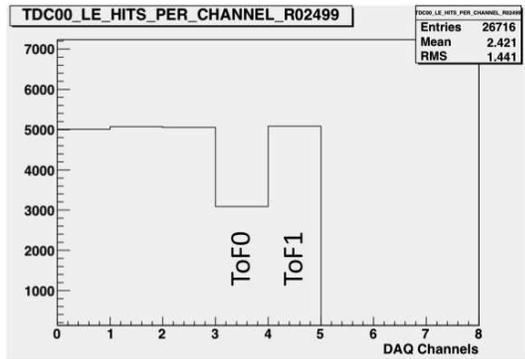}
   \caption{Online monitoring showing hits per channel in the data aquisition for negative polarity. Each channel is a scalar
	readout of either an individual detector or the number of triggers. Channels 4 and 5 show the number of recorded hits
	in TOF0 and TOF1 respectively. R02499 refers to run number 2499 conducted on July 26, 2010.}
   \label{hitsneg}
\end{figure}

\begin{figure}[htb]
   \centering
   \includegraphics*[width=70mm]{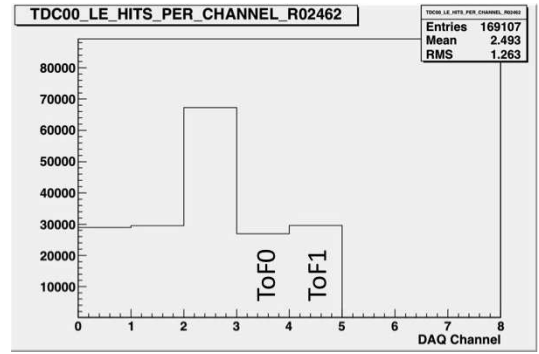}
   \caption{Online monitoring showing hits per channel in the data aquisition for positive polarity. Each channel is a scalar
	readout of either an individual detector or the number of triggers. Channels 4 and 5 show the number of recorded hits
	in TOF0 and TOF1 respectively. R02499 refers to run number 2462 conducted on July 21, 2010.}
   \label{hitspos}
\end{figure}

Data taking during the summer of 2010 began in May and ended in late August. With the major cooling channel components
not yet installed in the experimental hall, the main focus of this program was to characterize and measure the performance
of the upstream components of the beamline. During this period, a mismatch appeared in the hits registered in TOF0 and those
in the trigger station TOF1. It was observed for a negative beam in online monitoring plots (Fig. \ref{hitsneg}) in which TOF0 was showing about half as many
 particle hits as TOF1. Initially it was thought that an inefficiency had developed in TOF0. However, an inefficiency of
nearly 50\% seemed unlikely. For a positive beam, the mismatch was still observed but determined to be about 10\%, a
lower percentage of the overall particle rate (Fig. \ref{hitspos}). If there was an inefficiency in TOF0, the mismatch
should have been approximately the same percentage regardless of polarity. Because the beam is produced by protons on a
target, the production
rate of positive pions is higher than the rate for negative pions. Therefore, for a positive beam, the overall particle rate
at the TOFs is higher than for negative polarity. Looking again at the
mismatch, the absolute difference between the number of TOF triggers in both beam scenarios is about the same
(4-8 mismatches per spill) suggesting that the problem is not an inefficiency in TOF0.

Given that a detector inefficiency was unlikely, other possible causes for the extra hits in the TOF1 detector had to be
investigated. Cosmic ray particles are another potential source of this background. If they travel approximately
vertically, they will only pass through one detector, producing a trigger when hitting TOF1. However, cosmic ray particles
are independent of beam polarity, so they cannot be the only contribution to the mismatch.
As a result, the most likely candidate for the source of the mismatched hits is neutral particles from the beam. They will not ionize
the detector material, and will either pass through the detector without depositing any energy or collide with a nucleus producing
a signal but also stopping the neutral particle \cite{detector}.

\section{Analysis}
In order to confirm the existence of neutral particles in the beamline, several tests were performed.
First, data were taken with the target frame raised so that the target would not enter the ISIS proton
beam. Under these conditions the only triggers produced would be from cosmic ray particles. Counting the
number of hits in TOF1 without coincident hits in TOF0, we measure a mean of 0.16$\pm$0.02 per spill.
This rate is much lower than the apparent rate of mismatch under normal running conditions; therefore, more
than just cosmic rays are causing the mismatch.

Having shown that cosmic rays are only a small part of the background, beam conditions were varied again to look
for beam based sources of background. The target was lowered back into the beam, and the second dipole magnet
was turned off. In this configuration, no charged particles would be directed toward the TOFs but neutral particle paths
would not be affected. The mismatch rate under this condition was found to be very similar to the rate found
under normal running.

\begin{figure}[ht]
   \centering
   \includegraphics*[width=80mm]{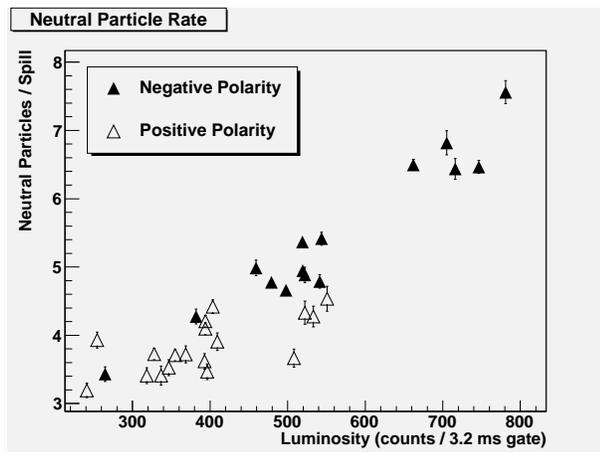}
   \caption{Mean number of neutral particles for various luminosities at reference magnet settings. Units of luminosity
	here are measured as coincident counts in the four scintillators per spill gate in the luminosity monitor.}
   \label{all_plots}
\end{figure}

As a final check, the beam stop, a thick plastic, steel, and concrete absorber located just upstream of TOF1,
was closed and the rate of mismatch was measured to be 0.73$\pm$0.2 per spill, again much smaller than the rate
under normal running conditions. These tests confirmed that the source of mismatch was coming from upstream of
the detectors. It could only be due to neutral particles that pass through TOF0 and stop in TOF1 causing a trigger without a hit in
TOF0. It appears that the particles originate in the target. If this is true, there should be a dependence on
the depth of the target in the proton beam, and the intensity of the beamloss generated. The
luminosity monitor, a set of four scintillators with a sensitive area of 4 cm$^2$ positioned 10 m from
the target, is used to measure the flux of particles coming from the target. This is directly related to the
depth of the target and the beamloss in the machine. The luminosity monitor output shows the number of particles coincident
 in all 4 scintillators during the 3.2 ms MICE data aquisition gate. Figure \ref{all_plots} shows the number of neutral particles per spill at various luminosities for both positive and negative polarities.
All data in Fig. \ref{all_plots} were taken with the MICE beamline magnet currents set to predefined daily reference settings.
The clear dependence on luminosity confirms that the neutral particles are beam based and produced at the target or in secondary
interactions.

When running with a positive beam, we employ a proton absorber upstream of the TOFs in order to limit triggers due to
proton contamination in the muon beam. This proton absorber consists of a set of plastic plates with different thicknesses
that can be lowered into the beam. A proton absorber thickness of 83 mm was used when taking data for the MICE reference
runs. The positive reference runs analyzed in Fig. \ref{all_plots} tend to show a lower neutral particle rate than negative
runs near the same luminosities. It is possible that we can use the proton absorbers to also limit the contamination of
neutral particles in the beam; however, more studies need to be done.

\section{conclusion and future work}
Although currently the rate of particles through the beamline is within the data taking capabilities of the DAQ,
as MICE moves to higher luminosities, we will approach its limits. We have established the existance of neutral
particles in the beam and have measured their effect on the TOFs and on the trigger rate. The proton absorbers
may reduce the number of neutral particles in the beam but more statistics will be needed to determine if the effect
is appreciable. We will continue to investigate methods to further filter them from the data. Examining individual
TOF PMT signals as digitized in flash ADCs may show a difference between muons passing through the detector and
neutral particles stopping in it. This would provide a way to identify neutral particles and remove them during beam analysis.
Understanding the neutral particle contamination and limiting its effect is especially important as our
luminosity increases toward the DAQ limits so that it does not reduce our percentage of useful data.


\begin{thebibliography}{9}   

\bibitem{MICE}
The International Muon Ionization Cooling Experiment (MICE) proposal to the Rutherford Appleton Laboratory, 10 January 2003 http://mice.iit.edu/mnp/MICE0021.pdf

\bibitem{coney}
L. Coney, ``MICE Overview'', DPF conf. proc. Detroit, MI, July 2009.

\bibitem{status}
MICE Status Report April 2010, MICE Collaboration, Y. Karadzhov \emph{et al.}, MICE note 288 {\footnotesize http://mice.iit.edu/micenotes/public/pdf/MICE0230/MICE0288.pdf}

\bibitem{tracker}
M. Ellis \emph{et al.}, ``The design, construction and performance of the MICE scintillating fibre trackers'', arXiv:1005.3491v2 [physics.ins-det], (2010).

\bibitem{tof}
R. Bertoni \emph{et al.} ``The design and commissioning of the MICE upstream time-of-flight system '', Nucl. Instr. Meth. A615 (2010) 14

\bibitem{detector}
D. Green, ``The Physics of Particle Detectors'', Cambridge, U.K., Cambridge University Press, 2000.

\end{thebibliography}
\end{document}